\documentclass[lettersize,journal]{IEEEtran}
\usepackage[T1]{fontenc}
\usepackage{amsmath,amsfonts}
\usepackage{algorithmic}
\usepackage{algorithm}
\usepackage{array}
\usepackage[caption=false,font=normalsize,labelfont=sf,textfont=sf]{subfig}
\usepackage{textcomp}
\usepackage{stfloats}
\usepackage{url}
\usepackage{verbatim}
\usepackage{graphicx}
\usepackage{cuted}
\usepackage{caption}
\usepackage{xcolor}
\usepackage{cite}
\usepackage{braket}

\usepackage{mathtools,graphicx} 

\usepackage{quantikz}

\usepackage[colorlinks=true,linkcolor=blue,citecolor=blue,urlcolor=blue]{hyperref}

\usepackage{libertinus}

\begin{document}


\title{Architectural Approaches to Fault-Tolerant Distributed Quantum Computing and Their Entanglement Overheads}

\author{
    \IEEEauthorblockN{
        Nitish Kumar Chandra\textsuperscript{†},
        Eneet Kaur\textsuperscript{‡},
        Kaushik P. Seshadreesan\textsuperscript{†}
    }

    \IEEEauthorblockA{
        \textsuperscript{†}Department of Informatics \& Networked Systems, School of Computing \& Information,\\
        University of Pittsburgh, Pittsburgh, PA 15260, USA\\
        Emails: nkc16@pitt.edu, kausesh@pitt.edu
    }

    \IEEEauthorblockA{
        \textsuperscript{‡}Cisco Quantum Lab, Los Angeles, CA 90404, USA\\
        Email: ekaur@cisco.com
    }
}


\markboth{Journal of \LaTeX\ Class Files,~Vol.~14, No.~8, August~2021}%
{Shell \MakeLowercase{\textit{et al.}}: A Sample Article Using IEEEtran.cls for IEEE Journals}


\maketitle

\begin{abstract}
Fault tolerant quantum computation over distributed quantum computing (DQC) platforms requires careful evaluation of resource requirements and noise thresholds. As quantum hardware advances toward modular and networked architectures, various fault tolerant DQC schemes have been proposed, which can be broadly categorized into three architectural types. Type 1 architectures consist of small quantum nodes connected via Greenberger–Horne–Zeilinger ($GHZ$) states, enabling nonlocal stabilizer measurements. Type 2 architectures distribute a large error correcting code block across multiple modules, with most stabilizer measurements remaining local, except for a small subset at patch boundaries that are performed using nonlocal CNOT gates. Type 3 architectures assign code blocks to distinct modules and can perform fault tolerant operations such as transversal gates, lattice surgery, and teleportation to implement logical operations between code blocks. Using the planar surface code and toric code as representative examples, we analyze how the resource requirements, particularly the number of Bell pairs and the average number of generation attempts, scale with increasing code distance across different architectural designs. This analysis provides valuable insights for identifying architectures well suited to fault tolerant distributed quantum computation under near term hardware and resource constraints.

\end{abstract}


\section{Introduction}

Fault-tolerant quantum computing (FTQC) offers a pathway toward executing large-scale quantum algorithms, even in the presence of noise and hardware imperfections~\cite{Campbell2017,régent2025awesomequantumcomputingexperiments,PRXQuantum.5.020101}. However, current quantum hardware is constrained by limited qubit counts, imperfect gate fidelities, and sparse connectivity. To overcome these limitations, \textit{distributed quantum computing} (DQC) has gained attention as a promising strategy for scaling up quantum systems~\cite{CALEFFI2024110672,BARRAL2025100747,PhysRevResearch.5.043302,10835728}. In this framework, multiple quantum processing units (QPUs) are linked via quantum communication channels, enabling non-local operations across physically separated devices. Achieving fault tolerance in distributed settings (FT-DQC) requires adapting or devising alternative methods to perform stabilizer measurements between qubits that are physically separated~\cite{LARASATI2025417}. Much of the existing work has focused on error-correcting codes such as the surface code and its variants, due to their high fault-tolerance thresholds and compatibility with current hardware~\cite{PhysRevA.86.032324}. However, distributed architectures present unique challenges, including the generation of high-fidelity entangled states, synchronization of operations across QPUs, and accounting for noise introduced by communication links.

Recent works have proposed various architectural frameworks for fault-tolerant distributed quantum computing (FT-DQC), each aiming to address the challenges of scaling quantum systems across spatially separated nodes~\cite{10.1116/5.0200190,chandra2025distributedrealizationcolorcodes,Ramette2024-ed,sutcliffe2025distributedquantumerrorcorrection,stack2025assessing,Nickerson2013,LARASATI2025417,clayton2025distributed,singh2024modular}. \textit{Type 1 architectures} are composed of quantum modules, each containing a small number of memory and communication qubits. These modules are optically connected, with communication qubits dedicated to generating entanglement pairs across nodes. Among the available memory qubits within a module, one is designated as the data qubit participating in the quantum error-correcting code, while the remaining memory qubits are used to store and process entangled states. Entangled states such as $GHZ$ states, once formed, are used to perform stabilizer measurements of the code. This architectural model is particularly suited to platforms such as nitrogen-vacancy (NV) centers in diamond, where the electron spin of the NV center functions as a communication qubit due to its desirable quantum properties and optical interface capabilities~\cite{Nickerson2013,10.1116/5.0200190}. The associated nuclear spins act as stable local memory qubits, capable of storing and processing quantum information.

The generation of high-fidelity $GHZ$ states across distributed nodes is a probabilistic process involving the fusion of multiple Bell pairs and the application of local Pauli corrections and measurements. Choosing an appropriate protocol requires balancing resource consumption with the desired fidelity. Several protocols such as \textit{Plain}, \textit{Basic, Medium, Refined, Expedient}, and \textit{Stringent} and several others have been developed to address this trade-off~\cite{Nickerson2013,10.1116/5.0200190,PhysRevX.4.041041,9292429}. These protocols differ significantly in both complexity and Bell pair requirements. For instance, the Plain protocol requires 3 Bell pairs, whereas the \textit{Refined} protocol consumes 40 Bell pairs to generate a 4-qubit $GHZ$ state, incorporating multiple rounds of purification to enhance fidelity.


\textit{Type 2 architectures} distribute large quantum error-correcting codes across multiple quantum modules, with inter-device operations implemented using entanglement-mediated non-local CNOT gates. These architectures have been shown to provide strong error suppression in boundary regions, where the presence of entanglement links introduces greater noise compared to the bulk of the code~\cite{Ramette2024-ed}. Such architectures are especially relevant for hardware platforms like superconducting qubits, which are typically fabricated as integrated circuits on separate chips. Given the area limitations of individual chips, inter-chip connectivity is essential for scaling up to large quantum error-correcting codes. For example, Ref.~\cite{Ramette2024-ed} proposes an architecture for linking two surface code patches using non-local CNOT gates. 
 Other implementations of Type 2 architectures have also utilized Floquet codes, distributed across multiple nodes~\cite{sutcliffe2025distributedquantumerrorcorrection}. Owing to their weight-2 stabilizer checks, distributed Floquet codes require only a few non-local stabilizer checks, since relatively few stabilizers span across nodes. In Ref.~\cite{chandra2025distributedrealizationcolorcodes}, a triangular color code is studied in a distributed configuration across four QPUs. In this configuration, the code’s weight-6 stabilizer checks require two ebits for performing each X- or Z-type syndrome measurement.

\textit{Type 3 architectures} describe a class of distributed quantum computing systems in which each node operates an entire logical code block that is used for computation, rather than being limited to memory storage~\cite{stack2025assessing}. Fault tolerant computations between nodes are enabled through non-local operations such as transversal gates, distributed lattice surgery procedures, or teleportation of logical states. One such architecture is proposed in Ref.~\cite{guinn2023codesignedsuperconductingarchitecturelattice}, where entangled Bell pairs, or ebits, are established between individual nodes and a central coordination module to facilitate inter-node operations. These entangled links are employed to carry out distributed lattice surgery between surface code blocks located on different quantum processors.  Later developments in Ref.~\cite{stack2025assessing} expand on this idea by demonstrating a circuit level simulation of a non-local CNOT gate implementation and a fault tolerant teleportation protocol on qLDPC Bivariate Bicycle (BB) and surface codes.

In this work, we discuss three distributed architectures (\emph{Type I} using GHZ mediated stabilizer measurements, \emph{Type II} with boundary connected patches, and \emph{Type III} considering teleportation and nonlocal CNOT). For \emph{Type I}, we derive an expression for the average number of entanglement attempts per syndrome round \(N_{\mathrm{round}}(d)\) as a function of entanglement generation probability, distillation success probability, error probability in the noise model, and code distance, and we compare the \textit{Plain, Basic, Medium,} and \textit{Refined} GHZ protocols. We quantify Bell pair costs in terms of expected entanglement generation attempts and their scaling with code distance across all three architectures.

In Sec.~\ref{bg}, we outline the background theory necessary to understand the architectural models and their operational primitives. Section~\ref{Archi} presents detailed descriptions of each architecture, highlighting the resource overheads and the expected number of Bell pair generation attempts needed for their implementation. We conclude with a discussion of the implications of our findings for scalable distributed quantum computing~\ref{conclusion}.

\section{Background Theory}
\label{bg}

In this section, we briefly review the background theory.

\subsection{Toric Code and Planar Surface Code}

These quantum error correction codes are defined on a square cellulation $G=(V,E,F)$ with vertices $V$, edges $E$, and faces $F$, where one physical qubit is placed on each edge $e\in E$. For every vertex $v\in V$ and face $f\in F$ the star and plaquette operators are defined as
\begin{equation}
A_v=\prod_{e\in\delta(v)} X_e,
\qquad
B_f=\prod_{e\in\partial f} Z_e,
\label{eq:star-plaquette}
\end{equation}
which commute and generate the stabilizer group $\mathcal{S}=\langle \{A_v\},\{B_f\}\rangle$. The code space is the simultaneous $+1$ eigenspace
\begin{equation}
\mathcal{C}=\big\{|\psi\rangle:\ A_v|\psi\rangle=|\psi\rangle,\ B_f|\psi\rangle=|\psi\rangle\ \text{for all } v,f\big\}.
\end{equation}
Logical Pauli operators are represented by elements of the normalizer $\mathcal{N}(\mathcal{S})$ that are not in $\mathcal{S}$. The distance is the minimum Hamming weight among nontrivial logicals,
\begin{equation}
d=\min_{L\in \mathcal{N}(\mathcal{S})\setminus\mathcal{S}} \operatorname{wt}(L).
\label{eq:distance}
\end{equation}
Logical classes correspond to noncontractible cycles on the primal and dual complexes \cite{KITAEV20032, 10.1063/1.1499754,PhysRevA.86.032324}.

\medskip
\noindent\textbf{Toric code.}
The toric code is defined by imposing periodic boundary conditions on an $L\times L$ lattice. The parameters on the square cellulation satisfy $|V|=|F|=L^{2}$ and $|E|=2L^{2}$, so $n=2L^{2}$ physical qubits are required. There are two global constraints $\prod_{v}A_{v}=I$ and $\prod_{f}B_{f}=I$, so the number of independent generators is $\operatorname{rank}(\mathcal{S})=2L^{2}-2$ and therefore the number of logical qubits is,
\begin{equation}
k=n-\operatorname{rank}(\mathcal{S})=2.
\end{equation}
Nontrivial logical operators are products of $Z$ or $X$ along noncontractible cycles on the primal or dual lattice \cite{KITAEV20032,10.1063/1.1499754}.

\medskip
\noindent\textbf{Planar surface code.}
If open boundaries are introduced in place of periodic ones, a planar surface code is obtained. Boundaries come in complementary types: rough boundaries terminate $Z$ strings and smooth boundaries terminate $X$ strings. In the rotated planar layout of odd distance $d$, the data qubit count is $n=d^{2}$ and a single logical qubit is encoded, $k=1$. A logical $Z_{L}$ is any product of $Z$ along a primal path connecting rough boundaries, and a logical $X_{L}$ is any product of $X$ along a dual path connecting smooth boundaries. The distance equals the minimal length of such boundary to boundary paths and equals $d$ for the rotated layout. The number of independent checks is $n-k=d^{2}-1$ \cite{PhysRevA.86.032324,marton2023coherent,PhysRevA.90.062320}.

\medskip

\noindent\textbf{Syndrome extraction.}
Stabilizer generators are measured using ancilla-mediated parity-check circuits, with corrections tracked in the Pauli frame. For a $Z$-type plaquette $B_f=\prod_{e\in\partial f} Z_e$, an ancilla $a_f$ is prepared in $\ket{0}$, entangled with each incident data qubit by CNOT gates with \emph{data as control} and \emph{ancilla as target} (one CNOT per edge in $\partial f$), and then measured in the $Z$ basis. The measurement outcome $m_{B_f}\in\{\pm1\}$ equals the eigenvalue of $B_f$.


For an $X$-type star $A_v=\prod_{e\in\delta(v)} X_e$, an ancilla $a_v$ is prepared in $\ket{+}$, entangled by CNOT gates with \emph{ancilla as control} and \emph{data as target} (one per edge in $\delta(v)$), and then measured in the $X$ basis. The outcome $m_{A_v}\in\{\pm1\}$ equals the eigenvalue of $A_v$. Up to local basis changes, these two circuits are equivalent to controlled-phase (CZ) implementations and constitute standard prepare–entangle–measure operations for toric and surface codes~\cite{10.1063/1.1499754,PhysRevA.86.032324}.


To suppress measurement errors, we repeat the same stabilizer checks over consecutive rounds indexed by \(t\).
Let \(m_g^{(t)}\in\{\pm 1\}\) denote the measured eigenvalue of generator \(g\in\{A_v,B_f\}\) at round \(t\).
We record a detection event whenever the sign flips between adjacent rounds, i.e.
\[
d_g^{(t)} \;:=\; \frac{1 - m_g^{(t-1)}\,m_g^{(t)}}{2} \in \{0,1\}.
\]
Space time decoding treats the collection of detection events as vertices in a three dimensional detection graph with edges across space and time.
A recovery is obtained by pairing these vertices with minimum total weight, typically using minimum weight perfect matching on that graph, which returns the state to the codespace while minimizing the probability of a logical error~\cite{higgott2025sparse}.

\subsection{Gate teleportation}

\begin{figure}[h!]
  \centering
  \includegraphics[width=\columnwidth]{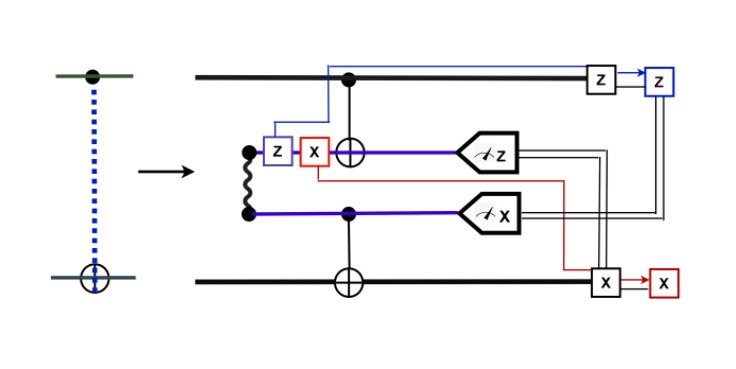} 
  \caption{Teleported non-local CNOT using a shared Bell pair. Local CNOTs and complementary-basis measurements implement a remote $\mathrm{CNOT}_{c\to t}$ with only classical feedforward across the link; noisy entanglement link errors propagate asymmetrically ($Z$ errors to the control, $X$ errors to the target).}
  \label{fig:teleported-cnot}
\end{figure}

Gate teleportation implements a non-local entangling gate between spatially separated qubits using a pre-shared Bell pair and classical feedforward (See Fig.~\ref{fig:teleported-cnot}). For a teleported $\mathrm{CNOT}_{c\to t}$ with control $c$ at node A and target $t$ at node B, one locally applies $\mathrm{CNOT}(c\!\to\!a)$ and $\mathrm{CNOT}(b\!\to\!t)$ where $a,b$ are the Bell-pair halves, measures $a$ in the $Z$ basis and $b$ in the $X$ basis, communicates the two outcomes, and then applies single-qubit Pauli corrections on $c$ and $t$ conditioned on those outcomes. Up to these conditioned corrections, the net operation equals the desired non-local CNOT; all entangling gates are local to each node and only the Bell pair and two classical bits are shared~\cite{CALEFFI2024110672,chandra2025distributedrealizationcolorcodes}.

Pauli noise propagates as shown in Fig.~\ref{fig:teleported-cnot}: a $Z$ (phase) error on the Bell pair propagates to the \emph{control} of the teleported CNOT, while an $X$ (bit-flip) error propagates to the \emph{target}. In distributed surface-code implementations this creates a seam whose error rate differs from the bulk. Stabilizer circuits that lie across the seam therefore use teleported CNOTs and noise models that assign larger or higher noise to the boundary or seam qubits\cite{Ramette2024-ed,chandra2025distributedrealizationcolorcodes}.

This quantum operation is quite useful in two settings. (i) \emph{Patching planar codes:} along the boundary between modules, non-local CNOTs in parity-check circuits can be implemented such that seam stabilizers can be measured while keeping all physical CNOT gates local to each module. (ii) \emph{Logical operations between distant code blocks:} distributed lattice surgery~\cite{guinn2023codesignedsuperconductingarchitecturelattice,ppng-vbqj} realizes logical measurements (merges/splits) through sequences of teleported interactions repeated for $O(d)$ rounds, and logical state transfer proceeds by preparing logical Bell pairs $|\Phi^+\rangle_L$ across nodes followed by a logical Bell measurement~\cite{stack2025assessing}.

\subsection{GHZ state preparation for distributed error correction}

In networked architectures, multi-party $GHZ$ states act as nonlocal ancillae for measuring weight-$w$ stabilizers across nodes: a $w$-qubit $GHZ$ is entangled to the $w$ data qubits, followed by local readout to obtain the joint parity, with no direct data--data interactions required. A standard way is to (i) create elementary Bell pairs between the involved nodes using heralded photonic schemes, (ii) \emph{fuse} Bell pairs to realize higher dimensional entangled states via local CNOTs and single-qubit measurements, and (iii) optionally \emph{distill} multiple Bell or $GHZ$ states to increase fidelity before being used in a stabilizer circuit~\cite{Nickerson2013,PhysRevX.4.041041}. For the surface/toric code, four-qubit $GHZ$ states are needed for star/plaquette checks on the square lattice.

Multiple $GHZ$ protocol families instantiate the fusion\,+\,distillation operations with different resource--quality trade-offs~\cite{Nickerson2013,PhysRevX.4.041041,9292429}. \emph{Plain} creates a four-qubit $GHZ$ by fusing three Bell pairs and uses no distillation, minimizing entanglement cost at the expense of fidelity.  Protocols such as \textit{Basic, Medium, Refined} uses two noisy $GHZ$ states and one is used to perform a 4-qubit parity projection onto the other.~\cite{PhysRevX.4.041041}. More recent optimization studies treat $GHZ$ generation as a search over Bell-pair fusions and nonlocal stabilizer measurements, identifying protocol “recipes” that maximize the final $GHZ$ fidelity for a given budget and quality of Bell pairs.~\cite{9292429,10.1116/5.0200190}.

\section{Architectural Designs for Fault-Tolerant DQC}\label{Archi}

Realizing fault-tolerant quantum computing over distributed architectures requires not only robust error correction codes but also scalable system designs that can operate under practical hardware constraints. In this section, we examine three representative architectural models: Type 1, Type 2, and Type 3. These models enable the implementation of fault-tolerant distributed quantum computing (DQC). To evaluate the feasibility and efficiency of these designs, we focus on their entanglement resource requirements as a function of code distance. Through our analysis, we assess the resource trade-offs associated with each architecture, providing insights into how architectural choices influence scalability and fault tolerance in distributed quantum systems.

\subsection{Type-I Architecture}
Type I architecture consists of connecting small modules with small numbers of memory and communication qubits. In this work, we consider the toric code in distributed settings where direct qubit connectivity across nodes with few qubits is not feasible. In such cases, $GHZ$ states can be used to perform stabilizer checks using nonlocal ancilla resources (see Fig.~\ref{fig:type1}).  In this distributed setting, one qubit of the $4$-qubit $GHZ$ state is assigned to each node involved in the stabilizer. Each $GHZ$ qubit interacts locally with the corresponding data qubit via a controlled gate, followed by a Pauli measurement. The combined measurement outcomes reveal the parity of the stabilizer operator, allowing the syndrome to be extracted nonlocally~\cite{10.1116/5.0200190}.

\subsection{GHZ State Preparation}
\label{subsec:GHZ-workflow}

\paragraph{Preparation Workflow}

Here, we discuss one of the methodologies to prepare a $GHZ$ state (See Fig.~\ref{fig:GHZ_state_gen}).
\vspace{3pt}
\begin{itemize}
  
  \item \textit{Entanglement Link Generation:} 
  We generate inter-cell Bell pairs over optical links between two nodes. A successful attempt yields one Bell pair on the two communication qubits. Given the entanglement link generation is probabilistic, we repeat the attempt until success and denote the per-attempt success probability by \(p_{\mathrm{link}}\).
\vspace{3pt}
  \item \textit{Entanglement Distillation:} 
 We perform entanglement distillation to obtain a Bell pair of increased fidelity from multiple lower-fidelity pairs, as specified by the protocol. In the two to one distillation protocol, we take two Bell pairs, apply bilateral CNOTs with the control qubits as controls and the target qubits as targets, and then measure the target qubits in the $Z$ basis. If the measurement outcomes satisfy the protocol’s keep condition (for example, the outcomes are identical), we retain the control pair; otherwise, both pairs are discarded and the procedure is repeated. Protocols such as \emph{Plain} omit this distillation step.

\vspace{3pt}

  \item \textit{Generating inter-cell $GHZ$ state:} Once a sufficient number of Bell pairs have been generated across the participating nodes, fusion operations (using local entangling gates and single-qubit Pauli-basis measurements) are applied to generate a $GHZ$ state. We denote \(n\) as the number of Bell pairs consumed by the chosen protocol to produce one $GHZ$ state.
  
\vspace{3pt}
       
\item \textit{Parity projection:} We prepare two noisy $GHZ$ states and perform a parity projection of one copy onto the other. Local CNOTs are applied, the projected copy is measured in the $Z$ basis, and the outcome is accepted only if the measurement parity is even. In that case, the unmeasured copy is retained with higher fidelity at the cost of consuming the second copy; otherwise, both copies are discarded and the procedure is repeated until acceptance.

\begin{figure}[h!]
  \centering
  \includegraphics[width=\columnwidth]{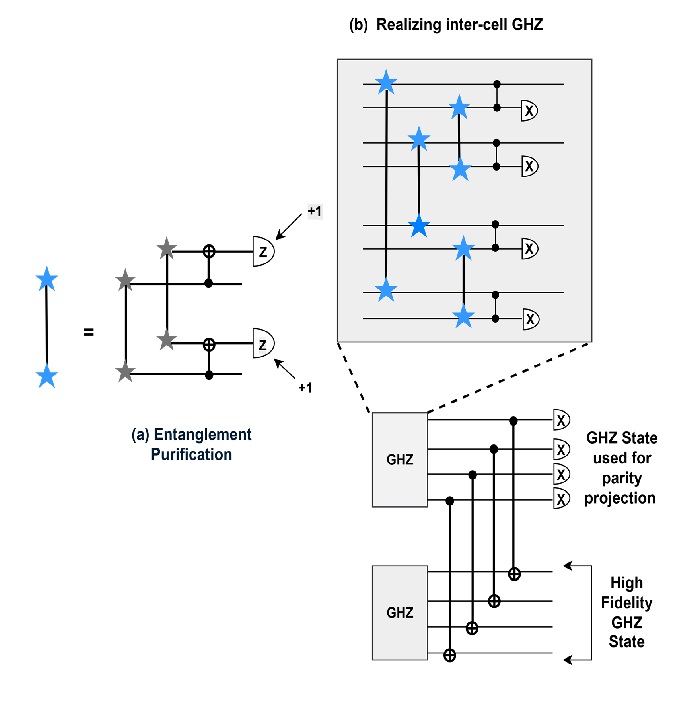} 
  \caption{\textit{Basic Protocol}: Entanglement purification and intercell $GHZ$ preparation with parity projection. 
(a) Two heralded Bell pairs (grey stars) are distilled using bilateral CNOTs and $Z$-basis measurements; when both outcomes are $+1$ as per  Extreme
Photon Loss (EPL) protocol~\cite{PhysRevLett.101.130502}, the remaining pair (blue star) has higher fidelity. 
(b) Purified intercell Bell pairs are combined by local entangling gates to form a distributed four-qubit $GHZ$. Two $GHZ$ copies are then used, with one serving as an ancilla to perform a four-qubit parity projection on the other; accepting even-parity outcomes yields a retained, higher-fidelity $GHZ$ resource for nonlocal stabilizer measurement.
}
  \label{fig:GHZ_state_gen}
\end{figure}

\end{itemize}
This workflow yields a single high-fidelity $GHZ$ from \emph{two} sets. Thus, a final $GHZ$ state
\emph{consumes \(2n\) two noisy Bell pairs in total} (before accounting for retries).

\vspace{5pt}

\paragraph{Protocols considered}
We consider four protocols whose circuit layout is shown in Fig.~\ref{fig:GHZ_state_gen}. In \textit{BASIC}, \textit{MEDIUM}, and \textit{REFINED}, two $GHZ$ copies are prepared and a four-qubit parity projection consumes one copy to increase fidelity~\cite{PhysRevX.4.041041}.

\begin{itemize}
  \item \textit{PLAIN (no distillation)}: Three Bell pairs are created and two fusion operations produce one four-qubit $GHZ$.
  \vspace{3pt}
  \item \textit{BASIC}: Each Bell pair used in $GHZ$ state generation is obtained by a two-to-one distillation step; thus $n = 8$ Bell pairs per $GHZ$ state. 
   \vspace{3pt}
  \item \textit{MEDIUM}: Each purified entangled pair (“blue star”) in Fig.~\ref{fig:GHZ_state_gen} uses four Bell pairs, requiring \(n=16\) Bell pairs per $GHZ$ state.
   \vspace{3pt}
  \item \textit{REFINED}: Each purified entangled pair uses ten Bell pairs, so \(n=40\) Bell pairs per $GHZ$ state. 
\end{itemize}

\subsection{Resource Overhead}

Here, we derive the expected number of
entanglement link generation attempts required to obtain
one high-fidelity GHZ state.
\paragraph{Model and Assumptions}
We denote \(n\) as the number of  Bell pairs required to generate one inter-cell $GHZ$ \emph{state} (the grey box in Fig.~\ref{fig:GHZ_state_gen}). A final high-fidelity $GHZ$ is obtained by preparing two noisy states and performing parity-projection of one state onto the other. Bell-pair generation, the \(2\!\to\!1\) distillation step, and the parity projection are modeled as independent Bernoulli trials with probabilities \(p_{\mathrm{link}}\), \(p_{\mathrm{distill}}\), and \(p_{\mathrm{parity}}\), respectively; for \textsc{Plain}, \(p_{\mathrm{distill}}=1\) implying no distillation has to be performed. We operate sequentially: successful intermediates are stored in memory and used in further operations.

\paragraph{Even-parity acceptance.}
Let \(S\in\{\pm1\}\) denote the measured parity (product of four \(X\) Pauli measurement outcomes on the projected copy). With single qubit-dependent depolarizing noise on the eight qubits, rates \(p_{A,i}\) and \(p_{B,i}\) on the two $GHZ$ copies (\(i=1,\dots,4\)), the parity moment and acceptance probability are,
\begin{equation}
\label{eq:parity-moment}
\mathbb{E}[S]=\prod_{i=1}^{4}\!\left(1-\tfrac{4}{3}p_{A,i}\right)\!\left(1-\tfrac{4}{3}p_{B,i}\right).
\end{equation}
\begin{equation}
\label{eq:parity-accept}
\begin{aligned}
p_{\mathrm{parity}}
&=\Pr(S=+1)=\tfrac{1}{2}\!\left[1+\mathbb{E}[S]\right] \\
&=\tfrac{1}{2}\!\Biggl[\,1+\prod_{i=1}^{4}
\Bigl(1-\tfrac{4}{3}p_{A,i}\Bigr)\Bigl(1-\tfrac{4}{3}p_{B,i}\Bigr)\,\Biggr].
\end{aligned}
\end{equation}

In the symmetric case \(p_{A,i}=p_{B,i}=p\),
\begin{equation}
\label{eq:parity-symmetric}
p_{\mathrm{parity}}=\tfrac{1}{2}\Bigl[1+\bigl(1-\tfrac{4}{3}p\bigr)^{8}\Bigr].
\end{equation}
We derive Equations~\eqref{eq:parity-moment}–\eqref{eq:parity-symmetric} in Appendix~\ref{app:GHZ-projection}.

\paragraph{Expected cost}
The expected number of attempts for one Bell pair is \(1/p_{\mathrm{link}}\). A successful distilled pair uses two entanglement pairs and succeeds with probability \(p_{\mathrm{distill}}\), so its average cost is \(2/(p_{\mathrm{link}}p_{\mathrm{distill}})\). One set contains \(n\) Bell pairs (equivalently \(n/2\) pairs); two sets therefore cost \(2n/(p_{\mathrm{link}}p_{\mathrm{distill}})\). Repeating the parity-projection until acceptance contributes a factor \(1/p_{\mathrm{parity}}\). With the shorthand
\[
B:=\frac{2}{p_{\mathrm{link}}\,p_{\mathrm{distill}}\,p_{\mathrm{parity}}}
\quad\text{and}\quad
B_{\textsc{plain}}:=\frac{2}{p_{\mathrm{link}}\,p_{\mathrm{parity}}}\,,
\]
the expected attempts per final $GHZ$ are
\begin{equation}
\label{eq:Rn-final}
\begin{aligned}
R(n) &= \frac{2n}{p_{\mathrm{link}}\,p_{\mathrm{distill}}\,p_{\mathrm{parity}}}=B\,n,\\
R_{\textsc{plain}}(n) &= B_{\textsc{plain}}\,n=\frac{2n}{p_{\mathrm{link}}\,p_{\mathrm{parity}}}.
\end{aligned}
\end{equation}

\paragraph{Protocol values.}
The value of \(n\) for the considered protocols are, \(\textsc{Plain}: n=3\), \(\textsc{Basic}: n=8\), \(\textsc{Medium}: n=16\), and \(\textsc{Refined}: n=40\). Substituting into \eqref{eq:Rn-final} yields
\begin{align}
R_{\textsc{plain}}   &= \frac{6}{p_{\mathrm{link}}\,p_{\mathrm{parity}}}, \\
R_{\textsc{basic}}   &= \frac{16}{p_{\mathrm{link}}\,p_{\mathrm{distill}}\,p_{\mathrm{parity}}}, \\
R_{\textsc{medium}}  &= \frac{32}{p_{\mathrm{link}}\,p_{\mathrm{distill}}\,p_{\mathrm{parity}}}, \\
R_{\textsc{refined}} &= \frac{80}{p_{\mathrm{link}}\,p_{\mathrm{distill}}\,p_{\mathrm{parity}}}.
\end{align}

The quantity \(R(n)\) in \eqref{eq:Rn-final} is the expected number of entanglement link generation attempts required to obtain \emph{one} accepted, high-fidelity $GHZ$ state. It accounts for all stochastic steps in the pipeline: creation of the \(n\) Bell pairs per set, the \(2\!\to\!1\) successful distillation (when used), preparation of two sets, and the parity-projection acceptance. Thus \(R(n)=B\,n\), where \(B=2/(p_{\mathrm{link}}\,p_{\mathrm{distill}}\,p_{\mathrm{parity}})\) aggregates the per-step success probabilities \(p_{\mathrm{link}}\), \(p_{\mathrm{distill}}\), and \(p_{\mathrm{parity}}\) (with \(p_{\mathrm{distill}}=1\) for \textsc{Plain}). Operationally, \(R(n)\) is the entanglement cost of one final $GHZ$; given an attempt rate \(\lambda\) (attempts per second), and the mean wall-clock time is \(R(n)/\lambda\).

\subsection{Resource overhead for stabilizer measurements in the toric code}

\begin{figure*}[t]
    \centering
    \includegraphics[width=0.80\textwidth]{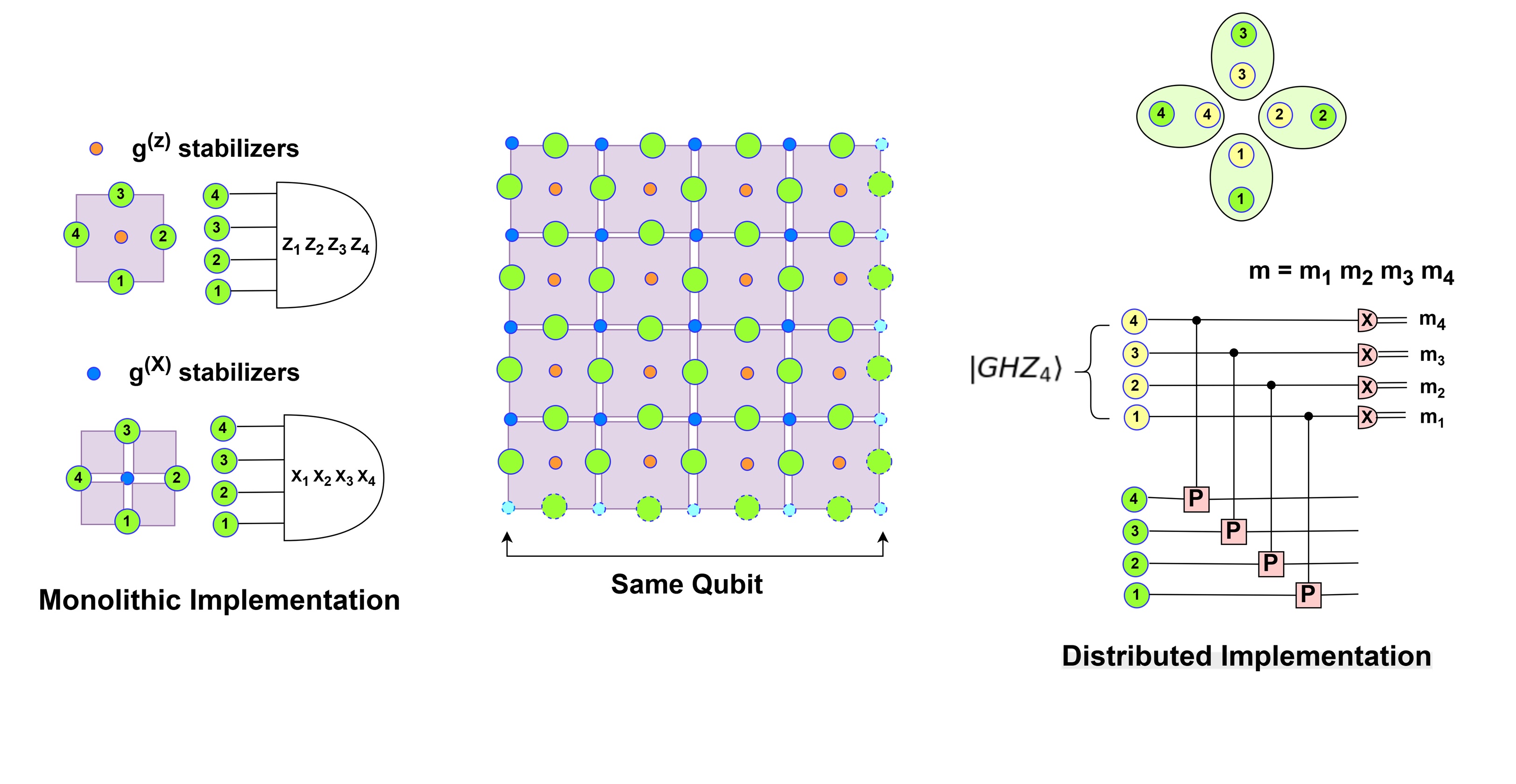}
    \caption{Schematic showing monolithic and distributed implementations of stabilizer measurements in the toric code with periodic boundary conditions. In the monolithic architecture, each stabilizer generator (either \(g^{(Z)}\) or \(g^{(X)}\)) is measured locally using an ancillary qubit that interacts with four neighboring data qubits. In the distributed implementation, stabilizer measurements are performed by preparing and distributing a $GHZ$ state across the involved nodes, followed by local Pauli measurements and classical communication to complete the nonlocal syndrome extraction. This figure is adapted from Ref.~\cite{10.1116/5.0200190}.}
    \label{fig:type1}
\end{figure*}

We consider the toric code on a two-dimensional periodic square lattice of distance \(d\), with physical qubits on edges so that there are \(2d^2\) qubits (See~Fig.~\ref{fig:type1}). The code encodes two logical qubits and has \(d^2\) plaquette (\(Z\)-type) and \(d^2\) star (\(X\)-type) stabilizers, each of weight four. Although two global constraints make one generator of each type redundant, for ease of comparison across distances and architectures we count the full set of \(2d^2\) stabilizers per round. In the distributed setting, each stabilizer measurement uses a four-qubit $GHZ$ state shared across the four incident nodes, so one round of syndrome extraction requires \(2d^2\) $GHZ$ states.

\begin{figure}[t]
  \centering
  \includegraphics[width=0.95\columnwidth]{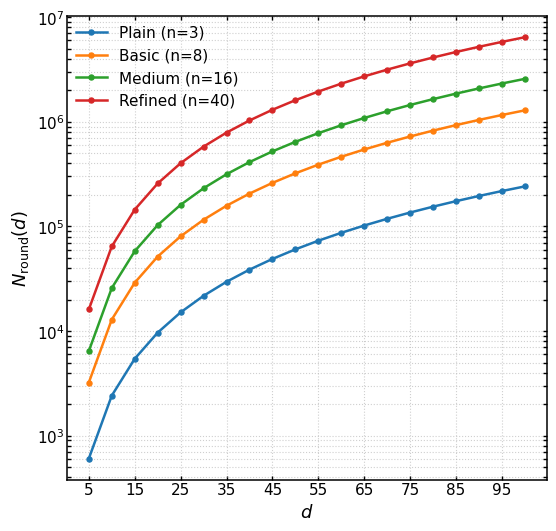}
  \caption{Expected entanglement link generation attempts per stabilizer round of each type ($X$ or $Z$) 
  $N_{\mathrm{round}}(d)$, versus code distance $d$ for the $GHZ$-mediated
  distributed setting. The parameters are $p_{\mathrm{link}}=0.5$ and $p=10^{-2}$ and $p_{distill} = 0.5$.
  }
  \label{fig:nround-vs-d}
\end{figure}

\begin{figure}[t]
  \centering
  \includegraphics[width=0.95\columnwidth]{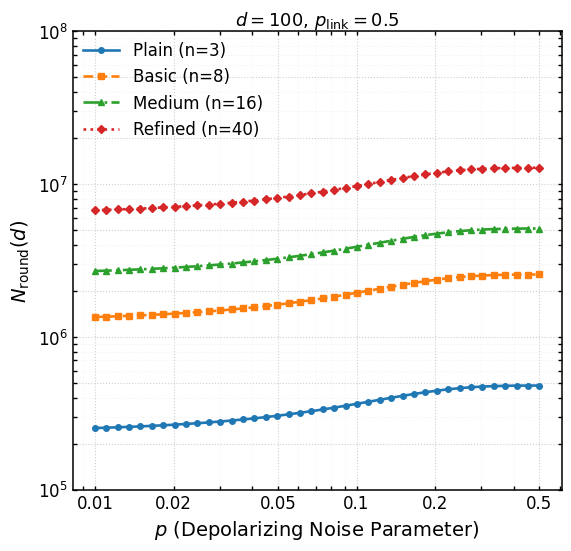}
  \caption{Expected entanglement link generation attempts per stabilizer round of each type ($X$ or $Z$) 
  $N_{\mathrm{round}}(d)$, versus depolarizing noise parameter $p$ for the $GHZ$-mediated
  distributed setting. The parameters are $p_{\mathrm{link}}=0.5$ and $d=100$ and $p_{distill} = 0.5$.
  }
  \label{fig:nround-vs-p}
\end{figure}

Let \(R(n)\) be the expected number of entanglement link generation attempts needed to obtain one \emph{accepted} $GHZ$  (Eq.~\eqref{eq:Rn-final}), where \(n\) is the number of Bell pairs per $GHZ$ set and \(p_{\mathrm{parity}}\) is derived in Appendix~\ref{app:GHZ-projection}. The entanglement cost (using entanglement link generation attempts) for each round
\noindent of ($X$ or $Z$) stabilizer measurements is:
\begin{equation}
\label{eq:Ntype}
N_{\mathrm{type}}(d)=d^{2}\,R(n).
\end{equation}
So,
\begin{align}
\label{eq:Nround}
N_{\mathrm{round}}(d)
&= \frac{4\,n\,d^{2}}{p_{\mathrm{link}}\,p_{\mathrm{distill}}\,p_{\mathrm{parity}}},\\[1pt]
\label{eq:NroundSym}
&= \frac{8\,n\,d^{2}}
       {p_{\mathrm{link}}\,p_{\mathrm{distill}}\!\left[\,1+\left(1-\tfrac{4}{3}p\right)^{8}\right]}.
\end{align}
We replace \(d^{2}\) by \(d^{2}\!-\!1\) in \eqref{eq:Ntype} when only independent generators are counted; the expression for \(N_{\mathrm{round}}\) follows analogously. The parameters \(n\), \(p_{\mathrm{distill}}\), and \(p_{\mathrm{parity}}\) encode the protocol choice: \textsc{Plain} has \(p_{\mathrm{distill}}=1\); a two-to-one distillation step has \(p_{\mathrm{distill}}<1\); the $GHZ$ parity-acceptance \(p_{\mathrm{parity}}\) is determined by the error model in Appendix~\ref{app:GHZ-projection}.\\

To illustrate one setting, we take the per attempt entanglement link success \(p_{\mathrm{link}}=0.5\), corresponding to the ideal, lossless, ancilla free two photon Barrett and Kok scheme in which a linear optical Bell state measurement succeeds with probability \(1/2\)~\cite{PhysRevA.71.060310}. For the Extreme Photon Loss (EPL) scheme, the distillation succeeds with probability \(p_{\mathrm{distill}}=\tfrac{1}{2}p_{R}^{2}\), where \(p_{R}\) is the parameter (equal to \(1/2\) for a perfectly generated Bell pair)~\cite{PhysRevLett.101.130502,PhysRevX.4.041041} for $R$ state defined as \( p_{R}\,|\Psi^{\pm}\rangle\langle\Psi^{\pm}| + (1-p_{R})\,|11\rangle\langle 11| \)
~\cite{PhysRevA.97.062333}. In architectures employing ancilla-assisted (“boosted”) Bell measurements, the BSM success can be raised toward \(3/4\)~\cite{PhysRevLett.113.140403}, so \(p_{\mathrm{link}}\) can, in principle, be extended to \(\sim 0.75\) (not used in our baseline). 

Figure~\ref{fig:nround-vs-d} shows expected entanglement link generation attempts per stabilizer round \(N_{\mathrm{round}}(d)\) versus distance \(d\) at fixed \(p\), \(p_{\mathrm{link}}\) and \(p_{distill}\). The growth is quadratic which reflects the ~\(d^{2}\) stabilizer checks per round in the toric code and the protocol-dependent constant of proportionality. Consequently, doubling \(d\) increases the required entanglement link attempts by roughly a factor of four, with the same vertical ordering of protocols (\textsc{Plain}~\(<\)~\textsc{Basic}~\(<\)~\textsc{Medium}~\(<\)~\textsc{Refined}) due to their respective \(n\) values.

Figure~\ref{fig:nround-vs-p} shows the expected entanglement link attempts per round, \(N_{\mathrm{round}}(d)\), versus the local depolarizing noise parameter\(p\) at fixed \(d\) and \(p_{\mathrm{link}}\). The curves increase monotonically with \(p\) because the GHZ parity-acceptance \(p_{\mathrm{parity}}(p)=\tfrac12\!\left[1+\bigl(1-\tfrac{4}{3}p\bigr)^{8}\right]\) decreases rapidly as \(p\) grows, and \(N_{\mathrm{round}}(d)\propto 1/p_{\mathrm{parity}}(p)\) [See Eq.~\eqref{eq:Nround}]. The four protocols appear as approximately vertical offsets on the log scale, set by their entanglement budget \(n\) (and \(p_{\mathrm{distill}}\)): \textsc{Plain} (smallest \(n\)) is lowest, while \textsc{Refined} (largest \(n\)) is highest. At small \(p\) the curves are relatively flat; for \(p\geq 0.05\) the eighth-power dependence becomes apparent with faster increase in the number of entanglement generation attempts.

Equation~\eqref{eq:Nround} can be optimized by improving the success probability of entanglement generation on each link and by choosing an appropriate distillation policy. With spatial or temporal multiplexing, if \(M\) independent attempts run in parallel within a time slot on a link, the effective success probability is \(p_{\mathrm{link}}^{\mathrm{eff}} = 1 - \bigl(1 - p_{\mathrm{link}}\bigr)^{M}\). Substituting \(p_{\mathrm{link}} \to p_{\mathrm{link}}^{\mathrm{eff}}\) in \eqref{eq:Nround} reduces \(N_{\mathrm{round}}(d)\) proportionally, subject to constraints on available communication qubits, reset times, and readout latencies~\cite{Dam_2017}. Other entanglement distillation protocols, such as BBPSSW and DEJMPS~\cite{PhysRevLett.77.2818,PhysRevLett.76.722}, have success probabilities that depend on the input state fidelity \(F\), denoted \(p_{\mathrm{distill}}(F)\), and can be considered and compared~\cite{PhysRevA.97.062333}.

\subsection{Type II Architectures}

\textit{Type II} quantum computing architectures enable scalable fault-tolerant computation by distributing large error correcting codes across multiple physical modules. Each module hosts a portion of the code, and these segments are integrated into a unified logical structure through stabilizer measurements that span module boundaries. These non-local checks are implemented using nonlocal CNOT gates facilitated by Bell pairs. A key feature of this architecture for patching planar surface codes is its higher tolerance to interface noise compared to bulk noise, due to inter-module boundaries being lower-dimensional than the code bulk~\cite{Ramette2024-ed}. This property allows the use of relatively noisy interconnects with minimal reduction in fault tolerance, offering a viable route to modular scaling without the additional cost of entanglement distillation.

\begin{figure*}[t]
    \centering
    \includegraphics[width=0.70\textwidth]{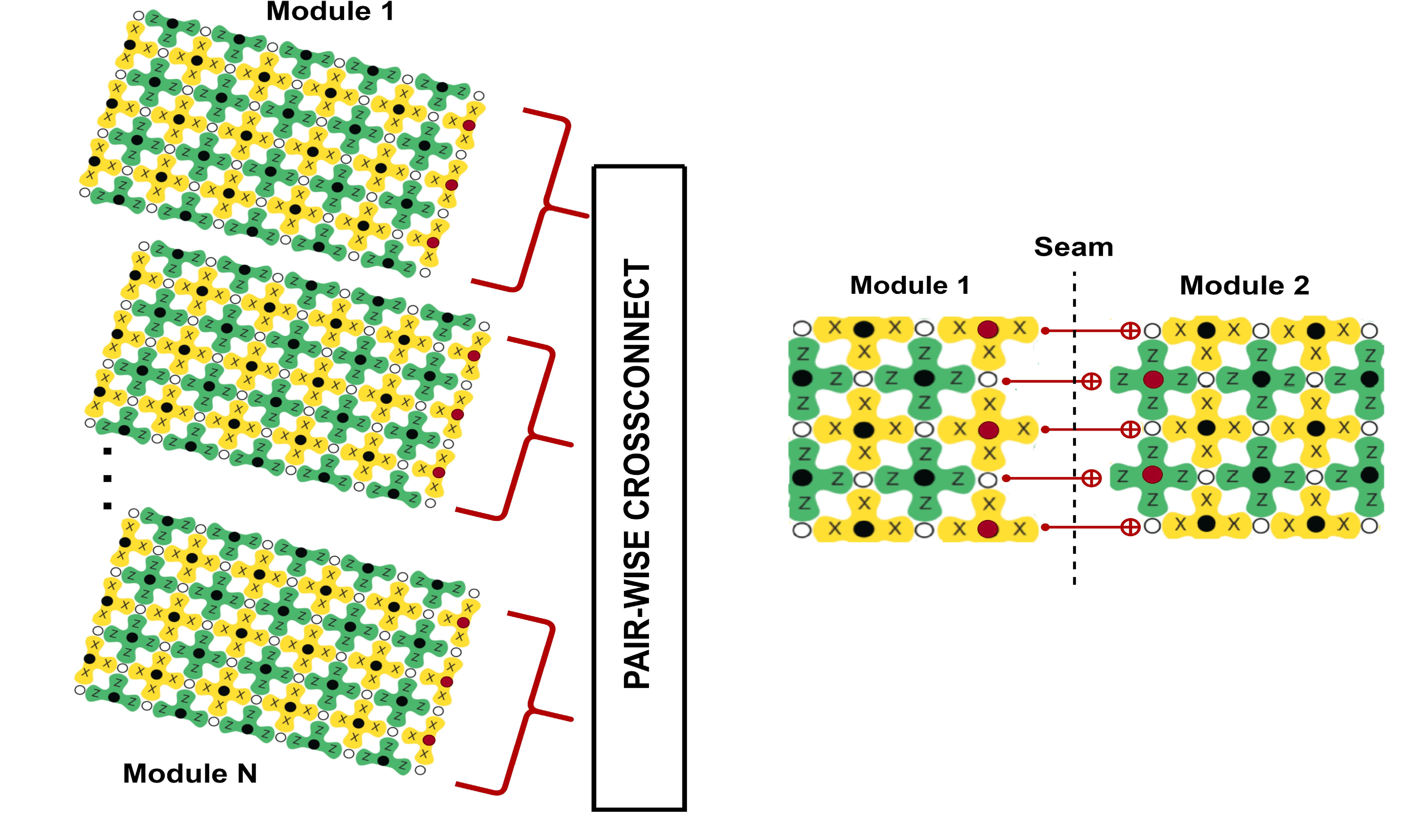}
    \caption{Left: Multiple surface code blocks are hosted on separate hardware modules, each connected via a reconfigurable pairwise cross-connect. This switching method enables connections between modules or allows them to be patched together. Right: A quantum operation between two surface code patches residing on different modules is mediated along a one-dimensional boundary or seam. Stabilizer measurements that cross the seam are performed using teleportation-based gates (shown in red). Data (open circles) and syndrome (filled circles) qubits located along this boundary are exposed to higher noise due to their participation in nonlocal entanglement generation. This figure is adapted from Refs.~\cite{PhysRevA.86.032324,Ramette2024-ed}.}

    \label{fig:type2}
\end{figure*}

An illustrative example of a Type II architecture involves stitching together two planar surface code patches along a shared boundary (See Fig.~\ref{fig:type2}). This strategy enables the construction of larger-distance logical qubit in modular systems and is well-suited to quantum platforms that support reconfigurable photonic or trapped-ion links for inter-module communication.

In a distance $d$ planar surface code, each boundary consists of $d$ data qubits and $d - 1$ syndrome qubits, yielding a total of $2d - 1$ physical qubits per boundary. When two such code patches are joined along a shared boundary, entanglement across the interface is required to facilitate distributed stabilizer measurements. A total of $2d - 1$ Bell pairs must be generated in every round to perform the X or Z type stabilizer checks across the interface. These per-round entanglement links represent a recurring resource cost necessary for executing syndrome extraction between modular code blocks.

\begin{figure}[t]
    \centering
    \includegraphics[width=0.50\textwidth]{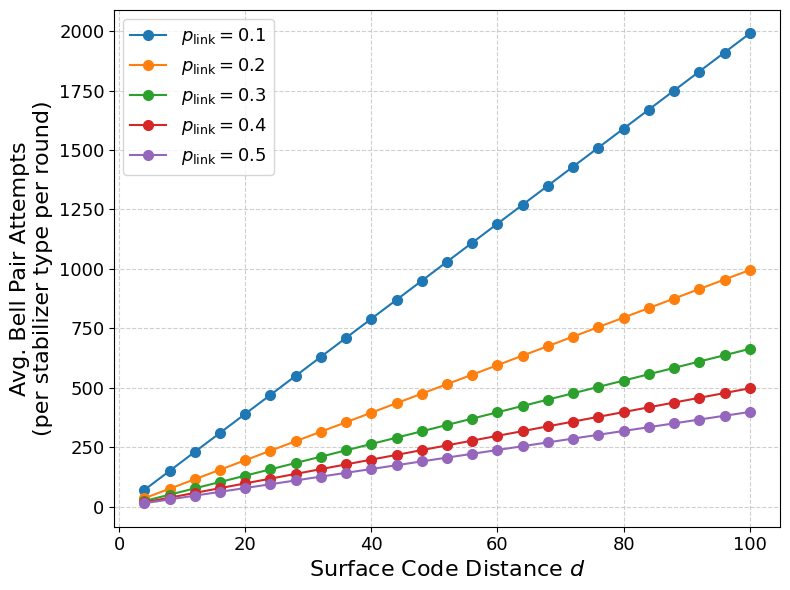}
    \caption{Expected Bell pair attempts per stabilizer type per round across a boundary between surface code patches, plotted for varying entanglement link generation success probabilities $p_{\mathrm{link}}$.}
    \label{fig:result2}
\end{figure}

Fig.~\ref{fig:result2} illustrates the expected number of Bell pair generation attempts required per stabilizer type (either X or Z) per round of syndrome extraction when two planar surface code patches are connected along a boundary. Given the probabilistic nature of entanglement generation, the average number of attempts scales as $\frac{2d - 1}{p_{\mathrm{link}}}$, where $p_{\mathrm{link}}$ is the success probability of a single Bell pair attempt. The plot shows this scaling for several values of $p_{\mathrm{link}}$ ranging from 0.1 to 0.5. As expected, higher success probabilities significantly reduce the entanglement cost per round, while the linear dependence on $d$ highlights the growing resource overhead at larger code distances.

\subsection{Type III Architectures}


\textit{Type 3} architectures support fault-tolerant quantum operations across spatially separated modules, with each module hosting one or more logical qubits based on the chosen error-correcting code. One approach to performing fault-tolerant computation with a topological code such as the planar surface code, where each code encodes a single logical qubit, is through lattice surgery. In this type of architecture, computation can be achieved either by performing distributed lattice surgery between separate code blocks using Bell pairs or by teleporting a logical state from one code block to another using a fault-tolerant teleportation protocol, followed by local computation~\cite{guinn2023codesignedsuperconductingarchitecturelattice,stack2025assessing}. We briefly outline how computation proceeds in each of these approaches and specifically analyze the number of entangled Bell pairs required to implement a CNOT gate between two logical qubits in the teleportation protocol, which enables the logical qubit to be transferred between modules.

\begin{figure*}[!t]
  \centering
  \includegraphics[width=0.65\textwidth]{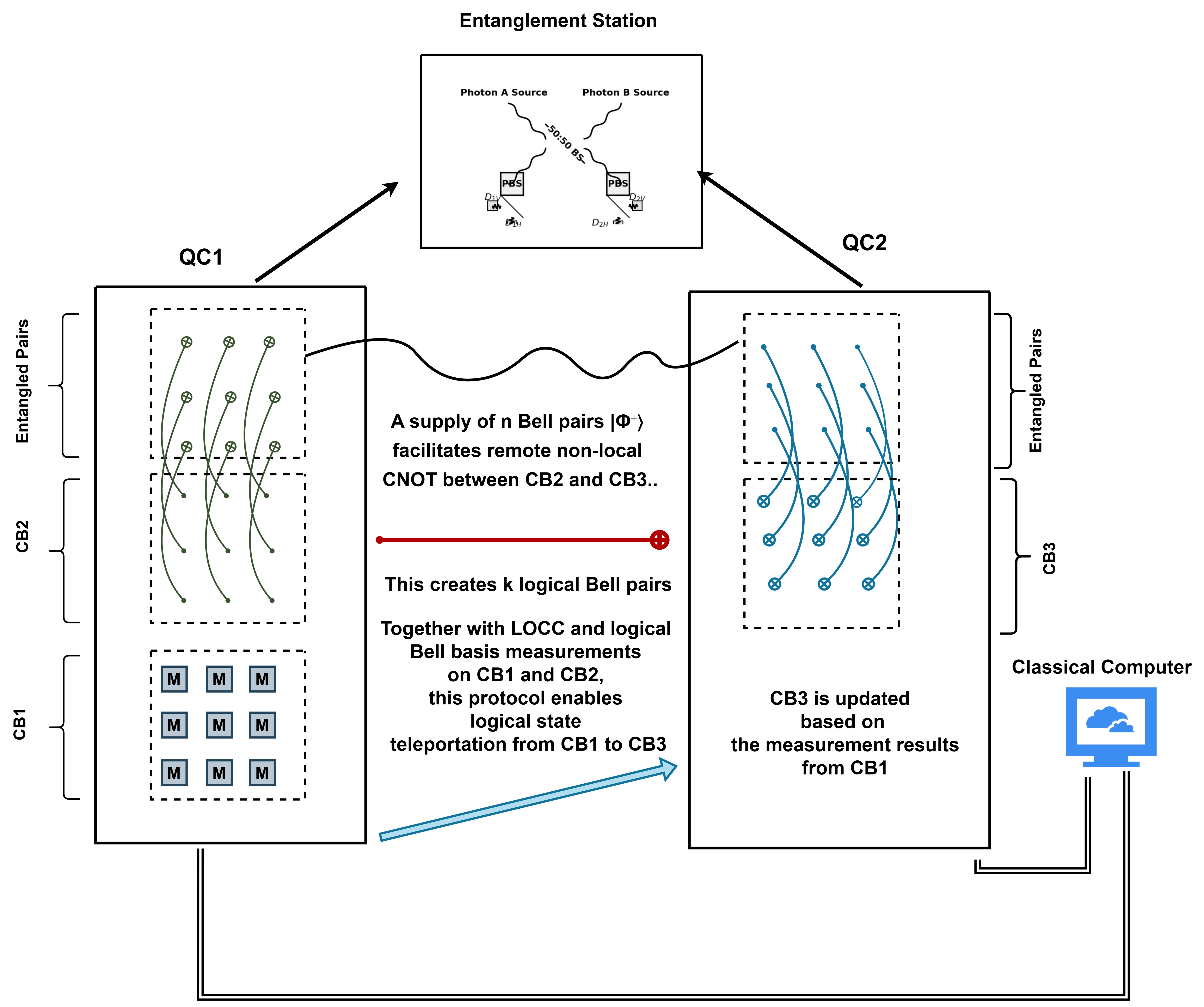}
  \caption{Depiction of a distributed quantum computing (DQC) system in which individual quantum processors are connected through a central entanglement station. The station comprises an optical switch, beam splitters, and detectors, and facilitates the creation of physical entangled Bell pairs between modules. On the left, quantum processor \texttt{QC1} contains code blocks \texttt{CB1} and \texttt{CB2}. \texttt{CB1} holds \texttt{k} logical qubits to be teleported, initialized in an arbitrary logical state \( |\psi\rangle \), while \texttt{CB2} is an ancilla patch initialized in \( |+\rangle_{L} \). On the right, \texttt{QC2} hosts \texttt{CB3}, another ancilla patch initialized in \( |0\rangle_{L} \). \texttt{k} logical Bell pairs are created between \texttt{CB2} and \texttt{CB3}. This shared entanglement, together with local operations and classical communication, enables a logical Bell measurement on \texttt{QC1} involving \texttt{CB1} and \texttt{CB2}. The outcome completes the teleportation of the logical state from \texttt{CB1} to \texttt{CB3}. This figure is adapted from Ref.~\cite{stack2025assessing}.}
  \label{fig:type3}
\end{figure*}

 Lattice surgery involves two fundamental operations: merging and splitting of code patches for measuring multiqubit operators~\cite{Horsman_2012}. For instance, the merge operation starts by initializing a column of physical qubits between two surface code patches in either the \( |0\rangle \) or \( |+\rangle \) state, depending on the boundary type. This initialization introduces new stabilizers that incorporate the intermediate qubits into the code. The merge is then carried out by measuring the set of stabilizers that include qubits from both code patches and the intermediate region. After \( d \) rounds of stabilizer measurements, if the intermediate qubits are initialized in the \( |+\rangle \) state, the resulting product of \( Z \)-type stabilizer outcomes yields a fault-tolerant measurement of \( Z_1 \otimes Z_2 \).



If the intermediate qubits are located within either of the modules, performing the merge requires nonlocal stabilizer measurements using Bell pairs, as described in the \textit{Type 2} architecture. The Bell pair requirement for a single round of merging two surface codes of distance \( d \) scales as \( \mathcal{O}(d) \). Since stabilizer measurements are repeated over \( d \) rounds, the total number of Bell pairs required scales as \( \mathcal{O}(d^2) \).

For Calderbank–Shor–Steane (CSS) codes such as surface codes, CNOT gates can be implemented transversally by applying CNOT gates between corresponding qubits in the two logical blocks. Fault-tolerant operations such as logical teleportation employ a combination of local CNOT gates, Bell pairs, ancilla patches, conditional Pauli corrections, and measurements to facilitate interactions between remote code blocks.


An illustrative application of this architecture is described in Ref.~\cite{stack2025assessing}, which involves teleporting a logical qubit from one module to another (see Fig.~\ref{fig:type3}). The goal is to teleport the logical state of surface code block CB1 on quantum processor QC1 to code block CB3 on quantum processor QC2. This requires initializing code block CB2 in the $|+\rangle_{L}$ state and CB3 in the logical state $|0\rangle_{L}$. Then, a logical Bell pair (k=1) is established between CB2 and CB3 using $n$ ebits, where $n$ corresponds to the number of physical qubits in the code, via a nonlocal CNOT gate (see Ref.~\cite{stack2025assessing}). Next, logical Bell measurements are performed on CB1 and CB2, and the outcomes determine the corrective operations to be applied to CB3. This enables successful logical state transfer from one module to another, after which computation can be performed locally using approaches such as lattice surgery or transversal gates.

Here, we analyze the resource cost in terms of Bell pairs required for teleporting a logical qubit from one module to another. This cost is determined by the structure of the quantum error-correcting code used to encode physical qubits into a logical qubit. For the distance $d$ rotated surface code, which has parameters of $[[d^2, 1, d]]$, there are $n = d^2$ physical data qubits that encode a single logical qubit. For performing a fault-tolerant non-local CNOT between two such code blocks requires one ebit pair for each physical data qubit, meaning a total of $n$ ebit pairs are needed. Consequently, the number of Bell pairs required to connect two distance $d$ surface code blocks is  $d^2$. The average number of attempts is given by $\frac{d^2}{p_{\mathrm{link}}}$. This quadratic relationship means that as the code distance is increased to improve error correction, the demand for high-fidelity entangled pairs between the quantum nodes grows rapidly, highlighting a significant challenge for scaling these distributed systems.


\begin{figure}[htbp]
    \centering
    \includegraphics[width=0.98\linewidth]{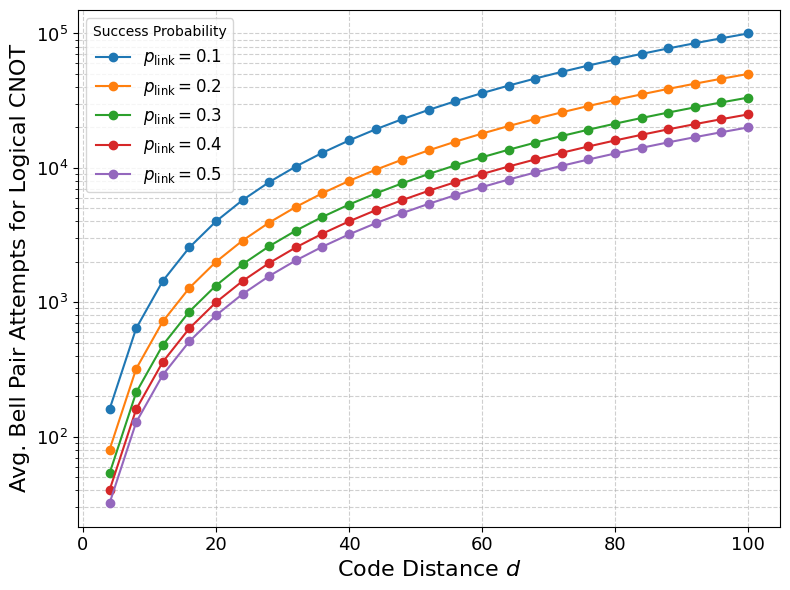}
    \caption{Average number of Bell pair generation attempts required to implement a non-local logical CNOT between two distance-$d$ surface code blocks, for various entanglement success probabilities $p_{\mathrm{link}}$. }
    \label{fig:logical-cnot-scaling}
\end{figure}

The Fig.~\ref{fig:logical-cnot-scaling} shows the expected number of Bell pair generation attempts required to implement a fault tolerant transversal CNOT between two distance $d$ surface code blocks, assuming probabilistic entanglement generation with success probability $p_{\mathrm{link}}$. The number of physical Bell pairs required for teleportation is the same as that required for performing a nonlocal CNOT gate between the two code blocks.

\section{Conclusion}\label{conclusion}
These distributed architectures offer distinct routes to do quantum computation, either for memory or for fault tolerant computation on modular platforms. \textit{Type I} performs stabilizer measurements using GHZ states, with an entanglement cost that scales as \(d^{2}\) and depends on the chosen GHZ generation protocol. For this architecture, we derived a closed form expression for the average number of entanglement attempts as a function of link success, distillation success, code distance, protocol choice, and a noise model for the circuit. Given the large entanglement demand, \textit{Type I} appears challenging with current technology. \textit{Type II} connects planar code patches along boundaries, requiring a fixed number of Bell pairs per syndrome round and yielding entanglement use that grows linearly with \(d\), which is favorable for quantum memory. \textit{Type III} enables logical operations between distant modules via logical CNOT gate and teleportation, with a \(\Theta(d^{2})\) Bell pair overhead per logical CNOT for the planar surface code. These classifications, organized by whether codes are used for memory or for logical operations, point to promising near term paths for scalable fault-tolerant distributed quantum computing and underscore the need for co-design across entanglement generation, code choice, hardware limits, and network protocols.

\section{Appendix: GHZ Projection}
\label{app:GHZ-projection}

In this appendix we compute the probability that a projection of one
$\mathrm{GHZ}_4$ state onto another $\mathrm{GHZ}_4$ state is \emph{accepted},
i.e., that the measured parity is even. The result is expressed in terms
of qubit error probabilities under an independent single-qubit depolarizing
noise model.

\subsection{Setting and Noise Model}

Let $A=(A_1,A_2,A_3,A_4)$ and $B=(B_1,B_2,B_3,B_4)$ be two four-qubit
registers, each prepared in the four-qubit $GHZ$ state
\[
\ket{\mathrm{GHZ}_4}=\frac{\ket{0000}+\ket{1111}}{\sqrt{2}}.
\]

We apply CNOT gates with control on $A_i$ and target on $B_i$ for $i=1,\dots,4$,
\begin{equation}
U=\bigotimes_{i=1}^{4}\mathrm{CNOT}_{A_i\to B_i},
\label{eq:U}
\end{equation}
and then perform $X-$ basis measurements on each $A_i$. We denote the outcomes by
$s_i\in\{+1,-1\}$ and define the projective parity as,
\begin{equation}
S:=\prod_{i=1}^{4}s_i .
\label{eq:S}
\end{equation}
We say the projection is \emph{accepted} when the parity is even, i.e., $S=+1$.

\paragraph{Noise model}
Immediately \emph{before} $U$, each qubit undergoes an independent
single-qubit depolarizing channel with parameter $p$,
\begin{equation}
\mathcal{D}_{p}(\rho)=(1-p)\rho+\tfrac{p}{3}\!\left(X\rho X+Y\rho Y+Z\rho Z\right).
\label{eq:depol}
\end{equation}
We allow different rates on the two registers and across wires, writing
$p_{A,i}$ for $A_i$ and $p_{B,i}$ for $B_i$.

\subsection{Required Results}

\paragraph{CNOT–Pauli conjugation (Heisenberg picture)}
Let $U=\mathrm{CNOT}_{A\to B}$. In the Heisenberg view we transform observables by
$P\mapsto UPU^\dagger$. The action on single–qubit Paulis is:
\begin{subequations}\label{eq:cnot}
\begin{align}
X_A &\mapsto X_A X_B, &\quad Z_A &\mapsto Z_A, \\
X_B &\mapsto X_B,     &\quad Z_B &\mapsto Z_A Z_B .
\end{align}
\end{subequations}

For a $\mathrm{CNOT}_{A\to B}$, an $X$ error on the control propagates to the target, and a $Z$ error on the target propagates to the control.

\vspace{2pt}

\paragraph{Measured parity in the Heisenberg picture}

We consider the parity observable on register \(A\), to be measured after applying \(U\):
\begin{equation}
M_A := \prod_{i=1}^{4} X_{A_i}.
\end{equation}
Since the single–qubit operators \(X_{A_i}\) commute, the product of the four outcomes
\(s_i\in\{\pm1\}\) equals the eigenvalue of \(M_A\) on the state after applying \(U\); i.e.,
\(S := \prod_{i=1}^{4} s_i\) is the measured parity.
In the Heisenberg picture, measuring \(M_A\) after applying \(U\) is equivalent to measuring its
conjugate \(U^\dagger M_A U\) before applying \(U\).
Using the Pauli conjugation rule for $U$, we obtain,
\begin{equation}
M_{AB} := U^\dagger M_A U = \prod_{i=1}^{4} (X_{A_i} X_{B_i}).
\label{eq:MAB}
\end{equation}
 Pauli errors that anticommute with \(M_{AB}\) (any local \(Z\) or \(Y\) on \(A_i\) or \(B_i\))
flip the recorded parity \(S\), whereas \(X\)-type errors commute and do not. Operationally,
multiplying the four \(X\) outcomes on \(A\) after applying \(U\) is the same as
measuring \(\prod_{i=1}^{4} X_{A_i} X_{B_i}\) before applying \(U\).

\vspace{5pt}


\paragraph{Products of independent Rademacher variables.}
Let \(A_1,\dots,A_m\) be independent Rademacher random variables (\(A_k\in\{\pm1\}\))~\cite{vershynin2018high}. Then
\begin{equation}
\mathbb{E}\!\left[\prod_{k=1}^{m} A_k\right]
=\prod_{k=1}^{m}\mathbb{E}[A_k].
\label{eq:rademacher}
\end{equation}

For any \(\{\pm1\}\) valued random variable \(B\),
\begin{equation}
\mathbb{E}[B]
=(+1)\Pr(B=+1)+(-1)\Pr(B=-1),
\end{equation}
\begin{equation}
\mathbb{E}[B]
=\Pr(B=+1)-\Pr(B=-1),
\end{equation}
\begin{equation}
\Pr(B=+1)+\Pr(B=-1)=1,
\end{equation}
hence
\begin{equation}
\Pr(B=+1)=\frac{1+\mathbb{E}[B]}{2},
\end{equation}
\begin{equation}
\Pr(B=-1)=\frac{1-\mathbb{E}[B]}{2}.
\end{equation}

\subsection{Parity Flips}

A local Pauli operator anticommutes with $X$ iff it has a $Z$ component. By
\eqref{eq:MAB}, a $Z$ component on $A_i$ or on $B_i$ flips the $i$th
contribution to the measured $X$-parity. So, we can define independent sign variables as,
\begin{subequations}\label{eq:zeta}
\begin{align}
\zeta_{A,i} &=
\begin{cases}
-1, & \text{if a $Z$ component is present on $A_i$,}\\
+1, & \text{otherwise.}
\end{cases}\label{eq:zetaA}\\[2pt]
\zeta_{B,i} &=
\begin{cases}
-1, & \text{if a $Z$ component is present on $B_i$,}\\
+1, & \text{otherwise.}
\end{cases}\label{eq:zetaB}
\end{align}
\end{subequations}

\begin{equation}
S=\prod_{i=1}^{4}\zeta_{A,i}\,\zeta_{B,i}.
\label{eq:Sproduct}
\end{equation}

\subsection{Mapping Depolarizing Noise to Parity Flips}

Using the depolarizing noise model as described in Eq.~\eqref{eq:depol} for each qubit, the error is $I$ with probability
$(1-p)$ and $X$, $Y$, or $Z$ with probability $p/3$ each. The $X$-parity
flips exactly when the local error has a $Z$ component, i.e., when it is
$Y$ or $Z$. Hence, for a $A_{i}$ or $B_{i}$,
\begin{subequations}\label{eq:eps-from-p}

\begin{align}
\Pr(\zeta=-1) &= \varepsilon = \tfrac{2}{3}p,\\
\Pr(\zeta=+1) &= 1-\varepsilon = 1-\tfrac{2}{3}p.
\end{align}
\end{subequations}
Thus,
\begin{equation}
\label{eq:Ezeta}
\begin{aligned}
\mathbb{E}[\zeta]
  &= (1-\varepsilon) - \varepsilon \\
  &= 1 - 2\varepsilon
   = 1 - \tfrac{4}{3}p .
\end{aligned}
\end{equation}
We can write the above equations in terms of qubit-dependent depolarizing rates,
\begin{subequations}\label{eq:meansites}
\begin{align}
\mathbb{E}[\zeta_{A,i}] &= 1 - \tfrac{4}{3}p_{A,i},\\
\mathbb{E}[\zeta_{B,i}] &= 1 - \tfrac{4}{3}p_{B,i}.
\end{align}
\end{subequations}

\subsection{Even-Parity Probability}

Taking expectations of both sides of Eq.~\eqref{eq:Sproduct}, and using
independence to factor the moment as in Eq.~\eqref{eq:rademacher}, together
with qubit-dependent depolarizing rates from Eq.~\eqref{eq:meansites}, yields

\begin{equation}
\mathbb{E}[S]
=\prod_{i=1}^{4}\bigl(1-\tfrac{4}{3}p_{A,i}\bigr)\bigl(1-\tfrac{4}{3}p_{B,i}\bigr).
\label{eq:ES}
\end{equation}
Since $S\in\{\pm1\}$, the probability of even parity is
\begin{align}
\Pr(S=+1)
&=\frac{1+\mathbb{E}[S]}{2}\nonumber\\[2pt]
&=\frac12\!\left[\,1+\prod_{i=1}^{4}
\bigl(1-\tfrac{4}{3}p_{A,i}\bigr)\bigl(1-\tfrac{4}{3}p_{B,i}\bigr)\right].
\label{eq:Peven-depol-general}
\end{align}

\paragraph{Symmetric special case.}
If all eight sites have the same depolarizing rate $p$, then
\begin{equation}
\Pr(\text{S=+1})=\tfrac12\Bigl[\,1+\bigl(1-\tfrac{4}{3}p\bigr)^{8}\Bigr].
\label{eq:Peven-depol-symm}
\end{equation}
For $p\ll1$,
\begin{equation}
\Pr(\text{S=+1})
=1-\tfrac{16}{3}p+\tfrac{224}{9}p^{2}+O(p^{3}).
\label{eq:series}
\end{equation}






\section{Acknowledgments}
 NKC and KPS gratefully acknowledge support from Cisco Systems, Inc..

\bibliographystyle{IEEEtran}  
\bibliography{ref}




\newpage

 




\vfill

\end{document}